\documentclass[prl,twocolumn,showpacs,preprintnumbers]{revtex4}
\usepackage{graphicx}
\usepackage{dcolumn}

\newcommand{\beq}{\begin{equation}}
\newcommand{\eeq}{\end{equation}}
\newcommand{\beqa}{\begin{eqnarray}}
\newcommand{\eeqa}{\end{eqnarray}}

\begin{document}

\title{Nodal/Antinodal Dichotomy and the Two Gaps\\
of a Superconducting Doped Mott Insulator}
\author{M. Civelli$^1$,
M. Capone$^2$,
A. Georges$^3$,
K. Haule$^4$,
O. Parcollet$^5$,
T. D. Stanescu$^6$
and G. Kotliar$^4$ }

\affiliation{$^1$ Theory Group, Institut Laue Langevin, Grenoble,
France}
\affiliation{$^2$ SMC, CNR-INFM, and Physics Department,
University of Rome ``La Sapienza'', Piazzale A. Moro 5, I-00185,
Rome, Italy and ISC-CNR, Via dei Taurini
19, I-00185, Rome, Italy}
\affiliation{$^3$ Centre
de Physique Th\'{e}orique, CNRS, Ecole Polytechnique, 91128 Palaiseau
Cedex, France}
\affiliation{$^4$ Physics Department
and Center for Materials Theory, Rutgers University, Piscataway
NJ USA}
\affiliation{$^5$ Service de Physique Th\'eorique,
CEA/DSM/SPhT-CNRS/SPM/URA 2306 CEA-Saclay, F-91191
Gif-sur-Yvette, France }
\affiliation{$^6$ Condensed Matter
Theory Center, Department of Physics, University of Maryland,
College Park, Maryland 20742-4111, USA }

\begin{abstract}
We study the superconducting state of the hole-doped
two-dimensional Hubbard model using Cellular Dynamical Mean Field
Theory, with the Lanczos method as impurity solver. In the
under-doped regime, we find a natural decomposition of the
one-particle (photoemission) energy-gap into two components. The
gap in the nodal regions, stemming from the anomalous
self-energy, decreases with decreasing doping. The antinodal gap
has an additional contribution from the normal component of the
self-energy, inherited from the normal-state pseudogap, and it
increases as the Mott insulating phase is approached.
\end{abstract}

\date{\today}
\maketitle

Superconductivity in strongly correlated materials such as the
high-$T_c$ cuprates has been the subject of intensive research for
more than twenty years (for a review see, e.g.,\cite{general04}).
From the theoretical side, low-energy descriptions in terms of
quasiparticles interacting with bosonic modes have been widely
studied starting from the weak correlation limit. A different
approach views the essence of the high-$T_c$ phenomenon as
deriving from doping with holes a Mott insulator \cite{anderson}.
The strong correlation viewpoint has not been yet developed into
a fully quantitative theory and whether the weak- and
strong-coupling pictures are qualitatively or only quantitatively
different is an important open issue.

The development of Dynamical Mean Field Theory (DMFT) and its
cluster extensions \cite{revmodmft} provides a new path to
investigate strongly correlated systems. These methods construct
a mean-field theory for Hubbard-like models using a cluster of
sites embedded in a self-consistent bath. Extensive
investigations have been carried out for intermediate
interaction-strength using the Dynamical Cluster Approximation on
large clusters \cite{jarrell-gen}. The strong coupling limit is
more difficult, as only small clusters can be employed. Many
groups however have identified interesting phenomena, such as the
competition between superconductivity and antiferromagnetism
\cite{dmft-ext}, the presence of a pseudogap (PG)
\cite{pg-bumsoo06}, the formation of Fermi arcs
\cite{bpk04,marce05,tudor-tudor06,berthod06} and the existence of
an avoided critical point \cite{haule-critical}. In this work we
use Cellular DMFT (CDMFT) to explore the energy gap in the
one-particle spectra of the superconducting state when
correlations are strong. The goal is to identify qualitative
aspects of the approach to the Mott transition in the light of
recent experimental studies on superconducting under-doped
cuprates \cite{tacon06,tanaka06,kondo07,Gomes_2007,earlier},
which report the presence of two distinct energy scales.

We consider the two-dimensional Hubbard Model:
\begin{equation}
\mathcal{H} = -\sum_{i,j,\sigma}t_{ij}\, c^{\dagger}_{i,\sigma}
c_{j,\sigma}  + U \sum_{i} n_{i\uparrow}n_{i\downarrow}
\label{hamiltonian}
\end{equation}
$c_{i,\sigma}$ destroys a $\sigma$-spin electron on site $i$,
$n_{i\sigma}= c^{\dagger}_{i\sigma} c_{i\sigma}$ is the number
operator and $t_{ii}\equiv \mu$ is the chemical potential. Only
next-neighbor $t$ and nearest-next-neighbor $t^{\prime}= -0.3t$
hoppings are considered. The on-site repulsion is set $U=12t$. We
implement CDMFT on a 2$\times$2 plaquette. Though this is the
minimal configuration allowing to study a d-wave superconducting
state, it already presents a rich physics and we think that its
deep understanding is an essential step to be accomplished before
challenging bigger clusters (hardly accessible by the
computational methods presently available). $\mathcal{H}$ is
mapped onto a 2$\times$2-cluster Anderson impurity model which is
solved using the Lanczos method \cite{krauth-caffarel}. The CDMFT
self-consistency condition \cite{venky05} is then enforced via the
Dyson relations $\hat\Sigma(i\omega_n)=
\hat\mathcal{G}^{-1}(i\omega_n)-\hat{G}^{-1}[\hat{\Sigma}](i\omega_n)$,
which determines the cluster self-energy $\hat{\Sigma}$. The hat
denotes 8$\times$8 matrices with cluster-site indices containing
both normal and anomalous components (Nambu notation).
$\hat\mathcal{G}$ is the "Weiss field" describing the bath, $\hat
G[\hat{\Sigma}]$ is the one-particle cluster Green's function
\cite{venky05} and $\omega_n= (2n+1)\pi/\beta$ the Matsubara
frequencies, with $\beta t=300$. The bath is described by 8
energy levels determined through a fit on the Matsubara axis
($0<\omega_{n}<2U$), which weights more the low frequencies
\cite{marce05}.

\begin{figure}[tb]
\begin{center}
\includegraphics[width=9.0cm,angle=-0]{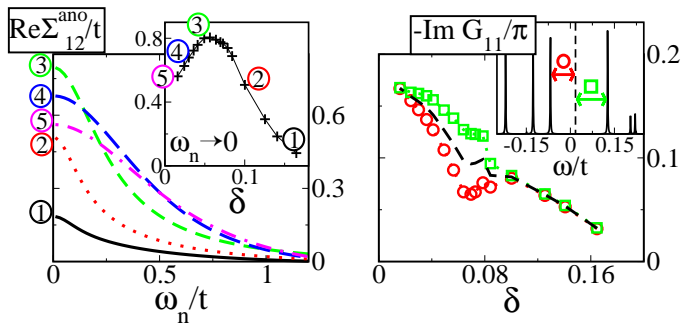}
\caption{(Color online). Left: Re$\Sigma^{\text{ano}}_{12}$ vs.
$\omega_n$. In the inset, the $\omega_n\to 0$ value as a function
of doping $\delta$. Right: The distance from the Fermi level
($\omega=0$) of the left (circle) and right (square) edge-peaks
in the local DOS $-\frac{1}{\pi}$Im$G_{11}$ (see inset) are
displayed as a function of $\delta$. The dashed line is the
average of the left and right values. In the inset $G_{11}$ is
shown for $\delta=0.06$ using a broadening $\eta\sim 7\times
10^{-3}t$ to display poles. } \label{Sig-ano-Gloc}
\end{center}
\end{figure}
Our main result is the presence of two energy-scales on the
under-doped side of the phase diagram. We first show that this
can be established directly from an analysis of quantities inside
the $2\times 2$ cluster, which are the output of the CDMFT
procedure. In the left panel of Fig. \ref{Sig-ano-Gloc} we display
the real part of the anomalous cluster self-energy
$\Sigma^{\text{ano}}$ on the Matsubara axis. Only the
nearest-neighbor component Re$\Sigma^{\text{ano}}_{12}(i\omega)$
is appreciably non-zero. The main observation is that at low
energy $\Sigma^{\text{ano}}_{12}(0)$ presents a non-monotonic
behavior with doping $\delta$, as emphasized in the inset. A first
characteristic energy-scale, measuring the superconducting
contribution to the one-particle energy-gap, can be defined as
$Z_{nod}\Sigma^{\text{ano}}_{12}(0)$, where $Z_{nod}$ is the
quasiparticle spectral weight at the nodal $k$-points, where
quasiparticles are well defined. As shown below, and as
physically expected, $Z_{nod}$ decreases as the doping is reduced
towards the Mott insulator. Hence,
$Z_{nod}\Sigma^{\text{ano}}_{12}(0)$ decreases too due to the
behavior of both $Z_{nod}$ and $\Sigma^{\text{ano}}_{12}(0)$. We
stress the sharp contrast of this result with resonating valence
bond mean-field (RVB-MF) theories \cite{patrick-rmp}, where
$Z_{nod}\Sigma^{\text{ano}}_{12}(0)$ corresponds to the spinon
pairing amplitude which is largest close to half-filling. In the
right panel of Fig. \ref{Sig-ano-Gloc} we show that there is
actually another energy-scale, which increases when the doping
level is reduced. This is revealed by looking at the local density
of state (DOS) $-\frac{1}{\pi} \hbox{Im}G_{11}$ in $\hat
G[\hat{\Sigma}]$. In the Lanczos-CDMFT the spectral function on
the real axis is obtained as a discrete set of poles (shown in
the inset), which are displayed by adding a small imaginary
broadening i$\eta$. We extract relevant energy scales by
measuring the distance from the Fermi level of the gap
edge-peaks. While for $\delta>0.08$ the spectrum is symmetric, an
asymmetry appears for $\delta<0.08$. The total energy gap (dashed
line in Fig. \ref{Sig-ano-Gloc}) grows with decreasing doping
$\delta$, as in RVB-MF theories.

In order to make contact with experimental observables it is
useful to obtain momentum-resolved quantities from the local
cluster quantities. For this we need a periodization procedure
restoring the translational invariance of the lattice. Several
schemes have been proposed \cite{revmodmft}. Building on previous
normal-state studies \cite{bpk04,tudor-tudor06} we use a mixed
scheme which is able to reconstruct the local cluster Green's
function (upon integrating over $k$ the lattice Green's function)
in the nodal and antinodal points better than uniform
periodization schemes. Our method is based on the idea that, when
the self-energies are regular the, most suitable choice is to
periodize the cluster self-energy via the formula
\begin{equation}
\Sigma_{\sigma}(k,\omega)= \,\frac{1}{N_{c}}\,
\sum_{\mu\nu} e^{-i k\mu}\,
\Sigma_{\mu\nu,\sigma}(\omega)
\,e^{ik\nu}
\label{periodization}
\end{equation}
($\mu, \nu$ label cluster sites). The anomalous self-energy
$\Sigma^{\text{ano}}$ and the normal self-energy
$\Sigma^{\text{nor}}$ {\it in the nodal regions}, where we expect
to find quasiparticles, are well behaved quantities, therefore we
extract them through formula (\ref{periodization}). In
particular, the anomalous self-energy acquires a standard
d$_{x^2-y^2}$-wave form: $\Sigma^{\text{ano}}(k,\omega)=
\,\Sigma^{\text{ano}}_{12}(\omega) \,\left( \cos k_{x}- \cos
k_{y} \right)$. On the other hand, when the self-energies develop
singularities, the cluster self-energy is not a good quantity to
be periodized. In Ref. \cite{tudor-tudor06}, it has been shown
that this takes place in the normal self-energy
$\Sigma^{\text{nor}}$ {\it in the antinodal regions}, when the
system approaches the Mott insulator. In this case, a more
suitable quantity to be periodized is the the irreducible
two-point cluster cumulant
$\mathcal{M}^{\text{nor}}_{\sigma}(\omega)= \left[(\omega+\mu)
\hat{1}- {\hat{\Sigma}^{\text{nor}}_{\sigma}} \right]^{-1}$,
which is a more local and regular quantity. In the antinodal
region, therefore, we can apply formula (\ref{periodization}) to
$\mathcal{M}^{\text{nor}}$, to obtain
$\mathcal{M}^{\text{nor}}(k,\omega)$ and finally extract the
normal lattice self-energy $\Sigma^{\text{nor}}(k,\omega)=
\omega+\mu- 1/\mathcal{M}^{\text{nor}}(k,\omega)$. The
$k$-dependent Green's function can be written as a matrix in
Nambu's space.
\begin{equation}
{\hat{G}}^{-1}_{k\sigma}(\omega)=  \left(
\begin{array} {cc}
\omega-t_{k}-\Sigma^{\text{nor}}_{\sigma}(k,\omega)
&-\Sigma^{\text{ano}}(k,\omega) \\
-\Sigma^{\text{ano}}(k,\omega) & \omega+t_{k}+
\Sigma^{\text{nor}}_{\sigma}(k,-\omega)^*
\end{array} \right)
\label{Gk-1}
\end{equation}
The imaginary part of the diagonal entry yields the spectral
function $A(k,\omega)$ measured in photoemission.

\begin{figure}[!!tb]
\begin{center}
\includegraphics[width=8.50cm,angle=-0]{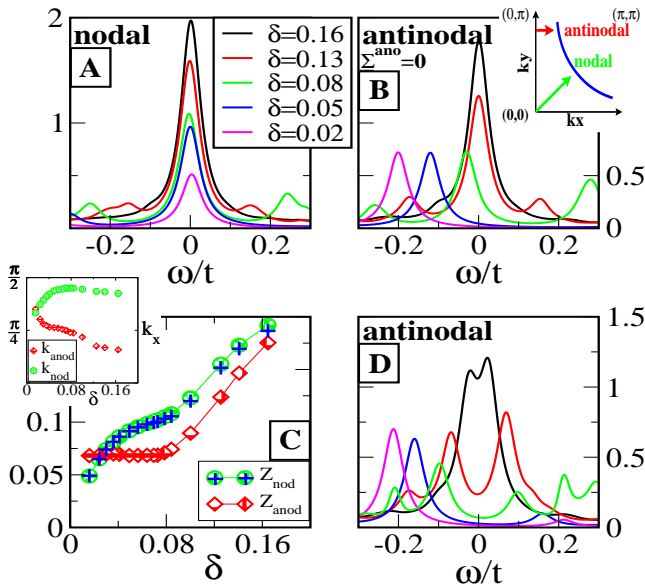}
\caption{(Color online) Spectral function $A(k,\omega)$ for
different $\delta$. Broadening $\eta=0.03 t$.
Panel A: nodal quasiparticle peak; Panel B, normal
component (set $\Sigma^{\text{ano}}=0$ in Eq. (\ref{Gk-1})) of
the antinodal quasiparticle peaks; Panel C, nodal and antinodal
quasiparticle weights. The inset shows the $k$-positions of the
nodal and antinodal points; Panel D, spectra at the antinodes.}
\label{Akw_ano-Z_ano}
\end{center}
\end{figure}

In order to compare our results with experiments, it is useful to
disentangle the normal and superconducting contributions to the
spectral gap. To this end, we first set $\Sigma^{\text{ano}}=0$
in Eq. (\ref{Gk-1}). The results are shown in Fig.
\ref{Akw_ano-Z_ano}. The $k$-points along the nodal and antinodal
directions are chosen as those where the highest peak is observed
in $A(k,\omega)$, as done, e.g., in Ref. \cite{tanaka06}. Their
actual values are shown in the inset of panel C of Fig.
\ref{Akw_ano-Z_ano}. Near the nodal point (panel A) a
quasiparticle peak is well defined at the Fermi level
($\omega=0$) and decreases by decreasing doping. In the antinodal
region (panel B), a quasiparticle peak is also found at the Fermi
level for $\delta>0.08$. For $\delta<0.08$, however, the spectral
weight shifts to negative energies signaling the opening of a PG,
whose size increases as $\delta\to 0$. The behaviour of the PG in
the superconducting solution smoothly connects to the PG
previously found in the normal state CDMFT study
\cite{tudor-tudor06}. The approach to the Mott transition is
characterized by a strong reduction in the area of the nodal
spectral peak $Z_{nod}$, which is plotted in panel C (green
circles). We also plot the area of the antinodal peak $Z_{anod}$,
which shows a constant value upon the opening of the PG ($\delta>
0.08$). In panel D, we restore $\Sigma^{\text{ano}}$, and examine
the actual superconducting solution. The superconducting gap
opens in the antinodal region (the nodal region is practically
unaffected). For $\delta>0.08$ the spectra are almost symmetric
around the Fermi level, as in a standard BCS d-wave
superconductor. In contrast, close to the Mott transition the PG,
which originates from the normal component, is superimposed to
the superconducting gap, resulting in asymmetric spectra. This
reveals the origin of the left/right asymmetry in the cluster DOS
discussed in Fig. \ref{Sig-ano-Gloc}.

\begin{figure}[!!t]
\begin{center}
\includegraphics[width=8.5cm,angle=-0]{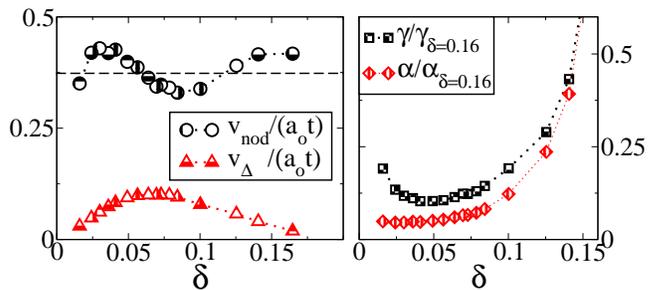}
\caption{(Color online) Left: $v_{nod}$ and $v_{\Delta}$ as a
function of doping $\delta$ ($a_{o}$ is the lattice spacing).
Right: low-frequency coefficients of local DOS $\gamma$ and of
the Raman B$_{2g}$ and superfluid density response $\alpha$,
renormalized by the value at $\delta=0.16$. } \label{vs}
\end{center}
\end{figure}

In the nodal region the quasiparticle peaks are well defined at
all dopings and we can expand the self-energies at low
frequencies. The quasiparticle residue $\left.(1-\partial_{\omega}
\hbox{Re}\Sigma_k(\omega))^{-1} \right|_{\omega=0}$ (blue crosses
in panel C of Fig. \ref{Akw_ano-Z_ano}) numerically coincides
with the area of the quasiparticle peak $Z_{nod}$. From Eq.
(\ref{Gk-1}), we get $A(k,\omega)\simeq Z_{nod}\, \delta\left(
\omega - \sqrt{ v_{nod}^{2} k^{2}_{\bot}+ v_{\Delta}^{2}
k^{2}_{\|}}\right)$, where  $v_{nod}= Z_{nod} |\nabla_{k}
\left(t_{k}-\Sigma^{\text{nor}}(k,0)\right)|$ and $v_{\Delta}=
\sqrt{2} Z_{nod} \Sigma^{\text{ano}}_{12}(0) |\sin k_{nod}|$ are
the normal and anomalous velocities respectively perpendicular
and parallel to the Fermi surface. $v_{\Delta}$ physically
expresses the superconducting energy-scale discussed in the left
panel of Fig. \ref{Sig-ano-Gloc}. We display them as a function of
doping $\delta$ on the left side of Fig. \ref{vs}. $v_{nod}$ does
not show a special trend for $\delta\to 0$ and it stays finite,
consistently with experiments \cite{Shen-Nature03}. The anomalous
velocity, $v_{\Delta} \ll v_{nod}$ presents a dome-like shape.
This behavior (confirmed by continuous time quantum Monte Carlo
(CTQMC) calculations~\cite{haule-ctqmc}) is in agreement with
recent experiments on under-doped cuprates showing that, contrary
to the antinodal gap, the nodal gap decreases by reducing doping
\cite{tacon06,tanaka06,kondo07}.

The low-energy behaviour of several physical observables in the
superconducting state is controlled by nodal-quasiparticle
properties and hence can be related to $v_{nod}$, $v_{\Delta}$
and $Z_{nod}$. Two specific ratios are particularly significant,
namely: $\gamma=Z_{nod}/(v_{nod}v_\Delta)$ and
$\alpha=Z_{nod}^{\,2}/(v_{nod}v_\Delta)$. The first one is
associated with the low-energy behaviour of the local DOS
measured in tunneling experiments: $N(\omega)=  \, \sum_{k}\,
A(k,\omega) \sim \gamma\,\omega$ (for $\omega\to 0$). Neglecting
vertex corrections \cite{tacon06}, the second one determines the
low-energy $B_{2g}$ Raman response function
$\chi^{\prime\prime}(\omega) \propto\alpha\,\omega$ and the
low-temperature ($T\to 0$) behaviour of the penetration depth
(superfluid density) $\rho_{s}(T)-\rho_{s}(0) \propto \, \alpha
T$. We display $\alpha$ and $\gamma$ in the right panel of Fig.
\ref{vs} as a function of $\delta$. $\alpha$ is monotonically
decreasing (see also CTQMC results \cite{haule-ctqmc}) and, on the
under-doped side $\delta < 0.08$, it saturates to a constant
value, in agreement with Raman spectroscopy \cite{tacon06} and
penetration depth measurements \cite{WHO}. Also $\gamma$ neatly
decreases in going from the over-doped to the under-doped side,
but it presents a weak upturn for low doping. The low-frequency
linear behavior of $N(\omega)$ is well established in scanning
tunneling experiments on the cuprates \cite{davis05}. However, it
is not currently possible to determine the absolute values of the
tunneling slope $\alpha$ from experiments, hence the behavior we
find is a theoretical prediction.

\begin{figure}[!!tb]
\begin{center}
\includegraphics[width=7cm,angle=-0]{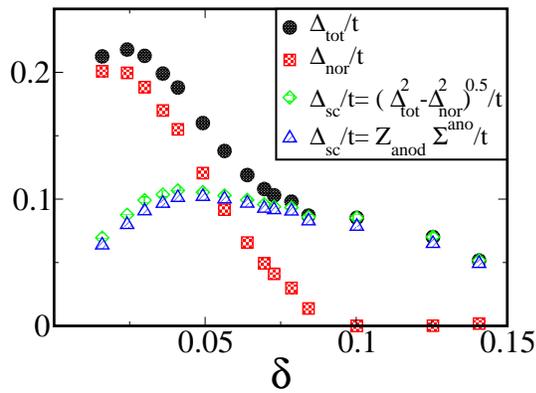}
\caption{(Color online) Antinodal energy gap $\Delta_{tot}$
(circles), obtained from the spectra of panel D in Fig.
\ref{Akw_ano-Z_ano}, as a function of doping $\delta$, and
decomposed in a normal contribution $\Delta_{nor}$ (squares),
obtained from panel B in Fig. \ref{Akw_ano-Z_ano}, and in a
superconducting contribution $\Delta_{sc}$ (diamonds). }
\label{n_gaps}
\end{center}
\end{figure}
We finally turn to the one-electron spectra in the antinodal
region, shown in Fig. \ref{n_gaps}, physically interpreting the
cluster energy-scales observed in Fig. \ref{Sig-ano-Gloc}. We
evaluate the antinodal gap in the superconducting state
$\Delta_{tot}$ by measuring the distance from the Fermi level
($\omega=0$) at which spectral peaks are located (panel D of Fig.
\ref{Akw_ano-Z_ano}). $\Delta_{tot}$ monotonically increases by
reducing doping, as observed in experiments. The data of panel B
in Fig. \ref{Akw_ano-Z_ano}, where $\Sigma^{\text{ano}}=0$, allow
us to extract the normal contribution $\Delta_{nor}$. We notice
that the peaks found there at negative frequency $\omega_{pg}$ do
not represent Landau quasiparticles in a strict sense, but we can
estimate the PG as $|\omega_{pg}|$. We also display the anomalous
contribution to the antinodal gap
$\Delta_{sc}=\sqrt{\Delta^{2}_{tot}-\Delta^{2}_{nor}}$, and find
that, within numerical precision, $\Delta_{sc}\simeq {\cal
Z}_{anod}|\Sigma^{\text{ano}}(k_{anod},\omega_{pg})| $. The
appearance of $\Delta_{nor}$ signs a downturn in $\Delta_{sc}$. We
interpret $\Delta_{tot}$ as the monotonically increasing
antinodal gap observed in cuprate superconductors, while the
superconducting gap $\Delta_{sc}$, detectable as the nodal-slope
$v_{\Delta}$ (Fig. \ref{vs}), is decreasing in approaching the
Mott transition.

The concept of two energy gaps with distinct doping dependence in
the cuprates has recently been brought into focus from an
analysis of Raman spectroscopy \cite{tacon06}, and photoemission
experiments \cite{tanaka06,kondo07}, which have revived
experimental and theoretical debate \cite{earlier}. Our
theoretical dynamical mean-field study of superconductivity near
the Mott transition establishes the remarkable coexistence of a
superconducting gap, stemming from the anomalous self-energy,
with a PG stemming from the normal self-energy. This is
reminiscent of slave-boson RVB-MF of the $t-J$ model
\cite{liu,patrick-rmp}, which uses order parameters defined on a
link and includes the possibility of pairing in {\it both} the
particle-particle and the particle-hole channels. Compared to the
self-energy of the RVB-MF, the CDMFT lattice-self-energy has
considerably stronger variations on the Fermi surface
\cite{tudor-tudor06} and additional frequency dependence, which
makes the electron states near the antinodes very incoherent even
in the superconducting state. Furthermore, in the RVB-MF theory
the anomalous self-energy monotonically increases by decreasing
doping, in contrast to our CDMFT results which reveal a second
energy scale associated with superconductivity, distinct from the
PG, which decreases with decreasing doping. Whether this feature
survives in larger clusters, representing a property of the real
ground-state, or it requires some further ingredient to be
stabilized against competing instabilities (above all
antiferromagnetism at low doping \cite{dmft-ext}) remains an
important open question addressed to future developments. We
think however that the assumption of a d-wave superconducting
ground-state is a reasonable starting point, and the importance
of our 2$\times$2-plaquette-CDMFT result stands in the natural
explanation it provides of the properties of under-doped cuprates.
\begin{acknowledgments}
We thank E.Kats, P.Nozi\`{e}res, P.Phillips, C.Castellani, A.-M.
Tremblay, B.Kyung, S.S. Kancharla, A. Sacuto and M. Le Tacon for
useful discussions. M.Ca. was supported by MIUR PRIN05 Prot.
200522492, G.K. by the NSF under Grant No. DMR 0528969.
\end{acknowledgments}


\end{document}